\documentclass[twocolumn,showpacs,showkeys,preprintnumbers,amsmath,amssymb]{revtex4}

\usepackage{graphicx}
\usepackage{url}
\usepackage{dcolumn}
\usepackage{bm}

\begin{document}

\title{Mutually-Antagonistic Interactions in Baseball Networks}

\author{Serguei Saavedra$^1$}
\email{s-saavedra@northwestern.edu}
\author{Scott Powers$^2$}
\author{Trent McCotter$^3$}
\author{Mason A. Porter$^{4,5}$}
\author{Peter J. Mucha$^{2,6}$}
\affiliation{$^1$Kellogg School of Management and NICO, Northwestern University, Evanston, Illinois, USA, 60208}
\affiliation{$^2$Carolina Center for Interdisciplinary Applied Mathematics, Department of Mathematics, University of North Carolina, Chapel Hill, NC 27599-3250, USA}
\affiliation{$^3$School of Law, University of North Carolina, Chapel Hill, NC 27599-3380, USA}
\affiliation{$^4$Oxford Centre for Industrial and Applied Mathematics, Mathematical Institute, University of Oxford, Oxford, OX1 3LB, UK}
\affiliation{$^5$CABDyN Complexity Centre, University of Oxford, Oxford, OX1 1HP, UK}
\affiliation{$^6$Institute for Advanced Materials, Nanoscience and Technology, University of North Carolina, Chapel Hill, NC 27599-3216, USA}

\date{\today}


\begin{abstract}
We formulate the head-to-head matchups between Major League Baseball pitchers and batters from 1954 to 2008 as a bipartite network of mutually-antagonistic interactions.  We consider both the full network and single-season networks, which exhibit interesting structural changes over time.  We find interesting structure in the network and examine their sensitivity to baseball's rule changes.  We then study a biased random walk on the matchup networks as a simple and transparent way to compare the performance of players who competed under different conditions and to include information about which particular players a given player has faced. We find that a player's position in the network does not correlate with his success in the random walker ranking but instead has a substantial effect on its sensitivity to changes in his own aggregate performance.
\end{abstract}

\pacs{64.60.aq, 02.50.-r, 05.40.Fb, 87.23.-n}

\keywords{bipartite networks,ranking systems, random walkers, competition dynamics}

\maketitle



\section{Introduction}\label{sec1}

The study of networks has experienced enormous growth in recent years, providing foundational insights into numerous complex systems ranging from protein interaction networks in biology to online friendship networks in the social sciences \cite{SIREV,Caldarelli,communities}. Research on ecological and organizational networks has provided a general framework to study the mechanisms that mediate the cooperation and competition dynamics between individuals \cite{Mayeco,Pascual,prey,Saa2,Saa,Bastolla}. In these networks, competitive interactions result from the indirect competition between members of different populations, who either compete for the same resources or are linked through consumer-resource relationships. However, data on mutually-antagonistic interactions---i.e., individuals who directly fight or compete against each other---have been more difficult to collect \cite{bacteria,parasites}. Mutually-antagonistic interactions also occur frequently in different social contexts such as sports. In the present paper, we consider head-to-head matchups between Major League Baseball (MLB) pitchers and batters: Pitchers benefit by ``defeating" batters, and vice versa. Using data from \url{retrosheet.org} \footnote{Retrosheet has pre-1954 data, but because that data is not currently complete, we do not extract pitcher--batter matchups from earlier years.}, we characterize the more than eight million MLB plate appearances from 1954 to 2008, considering full careers by examining head-to-head matchups over a multi-season (``career") network and single-season performances by constructing networks for individual seasons.

To compare the performance of players, MLB uses votes by professional journalists to recognize career achievement of players through induction into a Hall of Fame (HOF) and single-season performance through awards such as Most Valuable Player (MVP) and Cy Young (for pitching performance) \cite{BJ1}. Although the HOF purports to recognize the best players of all time, the selection of players to it is widely criticized by fans and pundits each year because of the lack of consistency when, e.g., comparing players from different eras, who play under fundamentally different conditions---in different ballparks, facing different players, etc.~\cite{Thorn, BJ2}. Such arguments come to the fore when attempting to draw comparisons between players elected to the HOF and others who did not make it. For instance, how can one tell whether Jim Rice (elected to the HOF in 2009) had a better career than Albert Belle (who dropped off the ballot because of low vote totals after only two years \footnote{Of course, the decisions of the HOF voters are influenced not only by perceptions of player performance in batter--pitcher matchups but also by perceptions of other factors such as defense, leadership/attitude/clubhouse presence, clutch performance, use of performance-enhancing drugs, etc. The current list of HOF members, players who will shortly become eligible for membership, and past voting results are available on the National Baseball Hall of Fame and Museum's website (\emph{www.baseballhalloffame.org})}? Does Bert Blyleven, who appeared on 62.7\% of the HOF ballots in 2009---short of the 75\% required for election---belong in the HOF?  Is Sandy Koufax, who played from 1955-1966 and is in the HOF, better than Pedro Martinez, who was still active during the 2008 season and who will presumably eventually be elected to the HOF? To address such questions, it is insufficient to rely purely on raw statistics; one must also consider quantitative mechanisms for comparison between athletes who played under different conditions. We take a first, simple step in this direction through the study of biased random walkers on these graphs \cite{notices,monthly}, allowing us to not only construct a quantitative, systematic, and transparent ranking methodology across different eras, but also to investigate the interplay between these dynamics and the underlying graph structure and to reveal key properties of mutually-antagonistic interactions that can potentially also be applied in other settings.  

While ``water-cooler" discussions about the HOF can often be fascinating, as indicated by the above paragraph, we stress that the primary goal of our paper is to investigate interesting features of the baseball \emph{networks} and the impact that network structure can have on rankings rather than on the rankings themselves.  While it is necessary to include some example rank orderings for the purpose of such a discussion, it is important to note that the rankings we show in the present paper must be taken with several grains of salt because our efforts at simplicity, which are crucial to highlighting the interplay between network structure and player rankings, require us to ignore essential contributing factors (some of which we will briefly discuss) that are necessary for any serious ranking of baseball players.

The rest of this paper is organized as follows. In Section \ref{sec2}, we define and characterize the mutually-antagonstic baseball networks and study the time evolution of various graph properties. In Section \ref{sec3}, we provide a description of the biased random walker dynamics that we employ as a ranking methodology across eras and for single-season networks. In Section \ref{sec4}, we study the interplay between the random walker dynamics and graph structure, paying special attention to the sensitivity of the player rankings. In Section \ref{sec5}, we conclude the paper and discuss a number of potential applications of our work.  We explain additional technical details in two appendices.

\section{Network Characterization and Evolution}\label{sec2}

\begin{figure*}
	\centering
	\begin{tabular}{c}
		\includegraphics[width=1.8in]{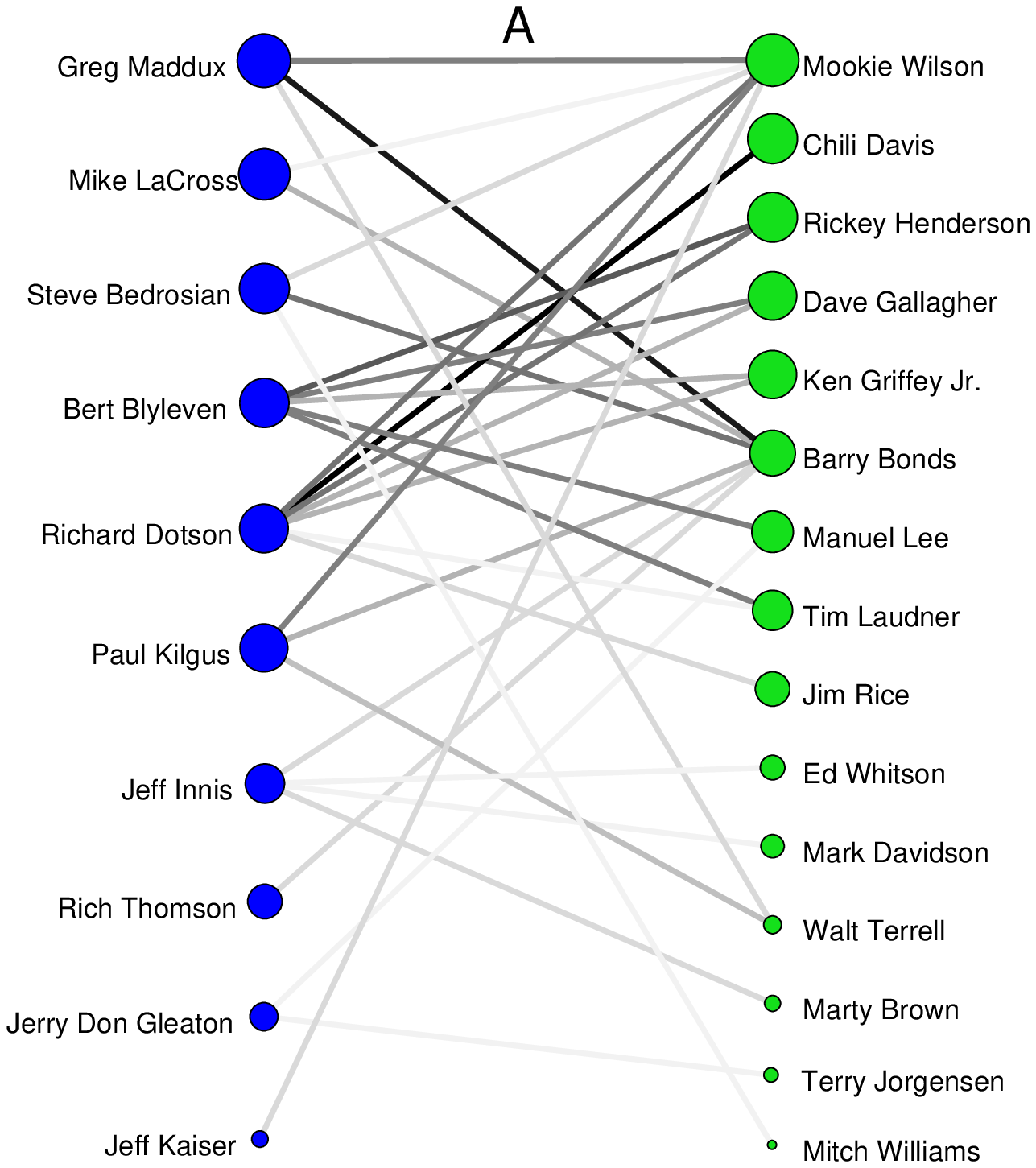}\\
		\includegraphics[width=3in]{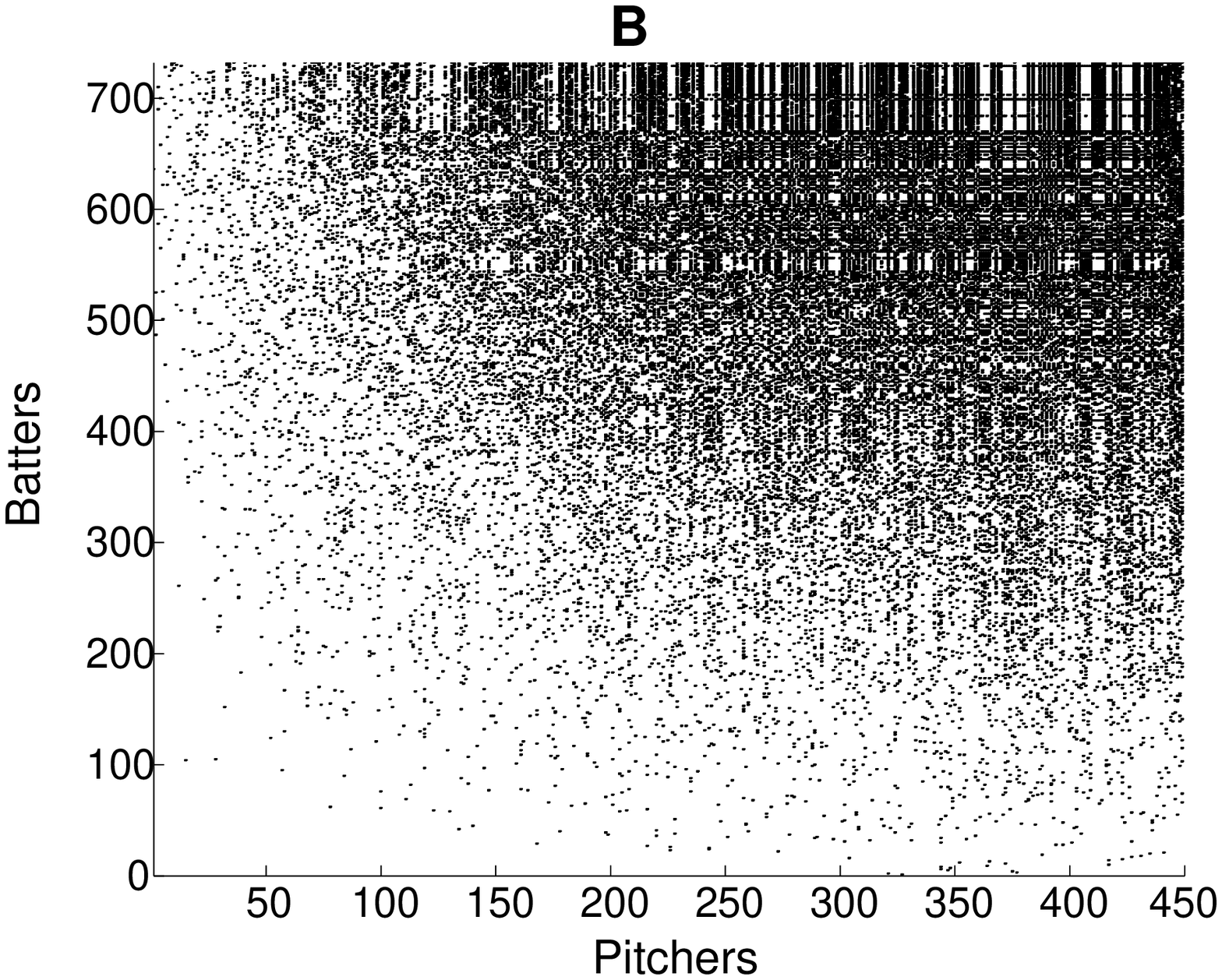}
		\end{tabular}
		\caption{Bipartite Baseball Networks. ({\bf A}) A subset of the bipartite interactions between pitchers (left column) and batters (right column) during the 1989 baseball season. The area of each circle is determined by the node degree (i.e., how many different opponents were faced).  Each line indicates that a given pitcher faced a given batter, and the darkness of each line is proportional to the number of plate appearances that occurred (i.e., the node strength). ({\bf B}) The matrix encoding the complete set of bipartite interactions from 1989, with pitchers (columns) and batters (rows) arranged from the lowest to the highest node degree.  An element of the matrix is black if that particular pitcher and batter faced each other and white if they did not. Observe the presence of a core of high-degree players that are heavily connected to each other (top right corner), an important presence of asymmetric interactions (i.e., high-degree players connected to low-degree players), and a dearth of connections between low-degree players (bottom left corner), which are all characteristics of nested networks \cite{nested}.  Some of the batters are actually pitchers (e.g., Mitch Williams), as National League pitchers (and, since 1997, also American League pitchers) have a chance to bat and face a small number of pitchers while at the plate.}
	\label{fig1}
\end{figure*}

We analyze baseball's mutually-antagonistic ecology by considering bipartite (two-mode) networks of head-to-head matchups between pitchers and batters. As shown in Fig.~\ref{fig1}, bipartite networks are formed using two disjoint sets of vertices, $P$ (pitchers) and $B$ (batters), and the requirement that every edge connect a vertex in $P$ to one in $B$ \cite{handbook,Saa,nested} (keeping the pitching and batting performances of pitchers as two separate nodes). We consider such interactions in terms of three different bipartite representations (with corresponding matrices): (1) The binary matchups $\mathbf{A}$ in which the element $A_{ij}$ equals $1$ if pitcher $i$ faced batter $j$ at any point and $0$ otherwise; (2) the weighted matchups $\mathbf{W}$ in which the element $W_{ij}$ equals the number of times that $i$ faced $j$; and (3) the weighted outcomes $\mathbf{M}$ in which the element $M_{ij}$ equals a ``score'' or performance index, which in the case of picther-batter matchups is determined using what are known in baseball as ``sabermetric" statistics (see the discussion below) \cite{Thorn,BJ2,dickson}, characterizing the results of all matchups between $i$ and $j$.  For each of these bipartite pitcher--batter networks, we also utilize corresponding square adjacency matrices:
\begin{equation*}
\hat{\mathbf{A}} = \begin{pmatrix} \mathbf{0} & \mathbf{A} \\
\mathbf{A}^T & \mathbf{0} \end{pmatrix} \,, \quad
\hat{\mathbf{W}} = \begin{pmatrix} \mathbf{0} & \mathbf{W} \\
\mathbf{W}^T & \mathbf{0} \end{pmatrix} \,, \quad
\hat{\mathbf{M}} = \begin{pmatrix} \mathbf{0} & -\mathbf{M} \\
	\mathbf{M}^T & \mathbf{0} \end{pmatrix} \,,
\end{equation*}
so that they are appropriately symmetric ($\hat{\mathbf{A}}$ and $\hat{\mathbf{W}}$) and anti-symmetric ($\hat{\mathbf{M}}$). We construct and analyze each of these representations for the single-season networks and the aggregate (career) network that contains all pitcher--batter interactions between 1954 and 2008.

\begin{figure*}
\begin{center}
\begin{tabular}{c}
\includegraphics[width=2.5in]{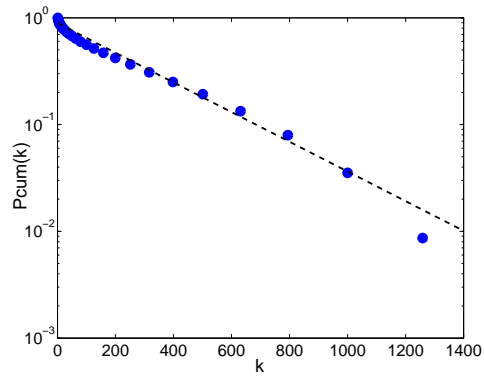}
\end{tabular}
\end{center}
\caption{[Color online] Cumulative Degree Distribution. Semi-log plot of the cumulative degree distribution $P_{cum}(k)$ for pitchers and batters in the career (1954--2008) network. The empirical data (dots) are arranged in logarithmic bins.
}
\label{figS1}
\end{figure*}

\begin{figure*}[p]
\begin{center}
\begin{tabular}{cc}
\includegraphics[width=2.5in]{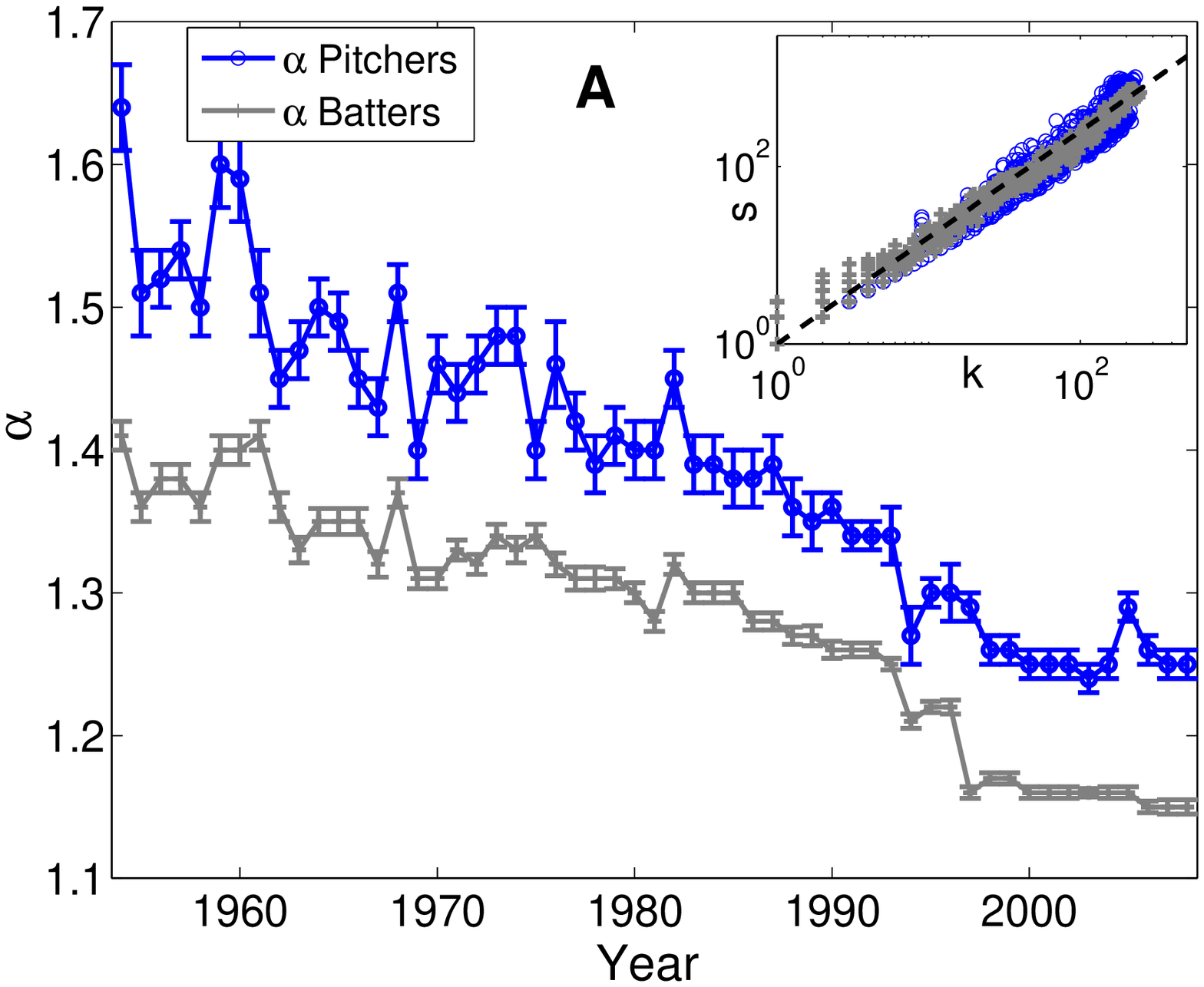}&\includegraphics[width=2.5in]{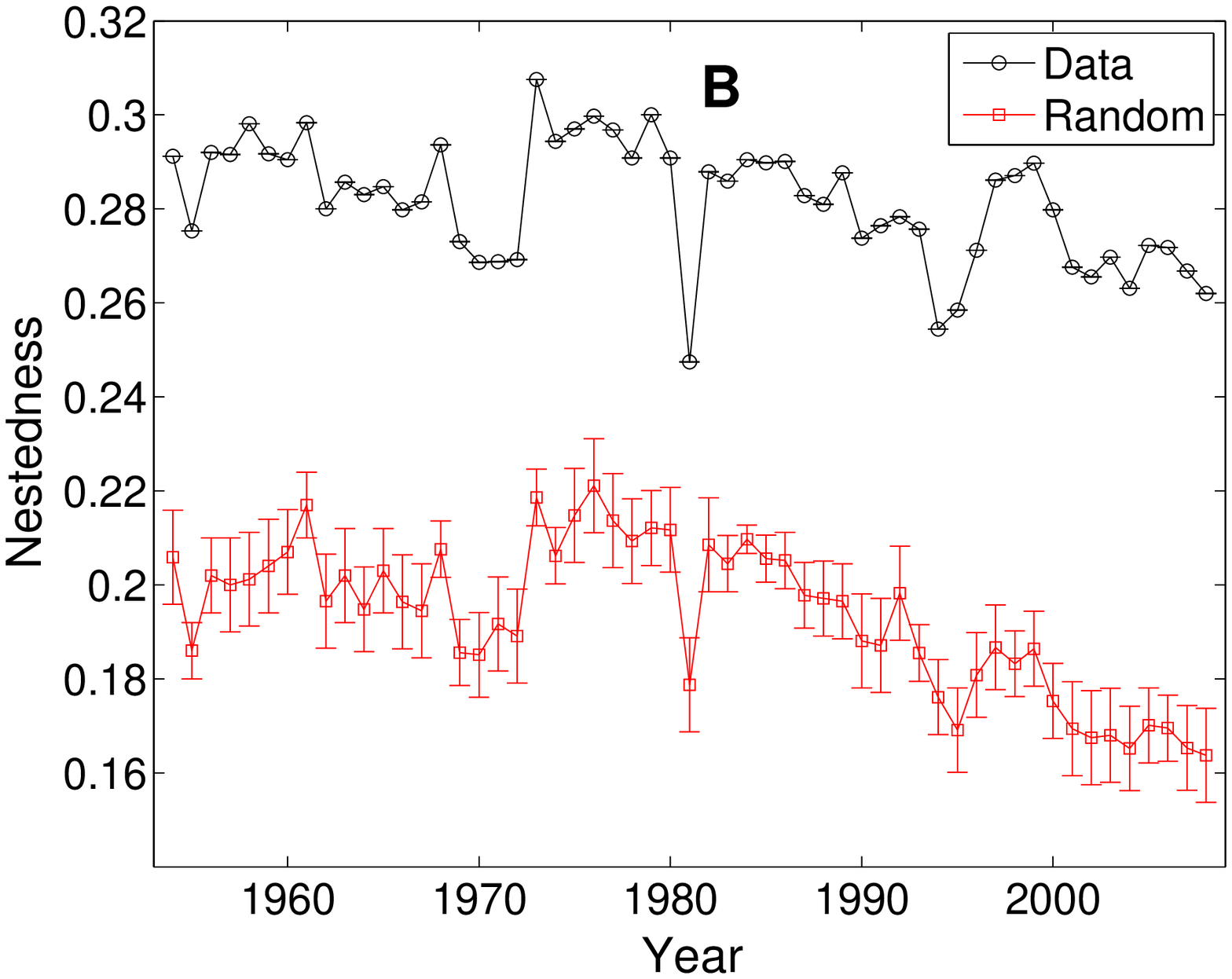} \\
\includegraphics[width=2.5in]{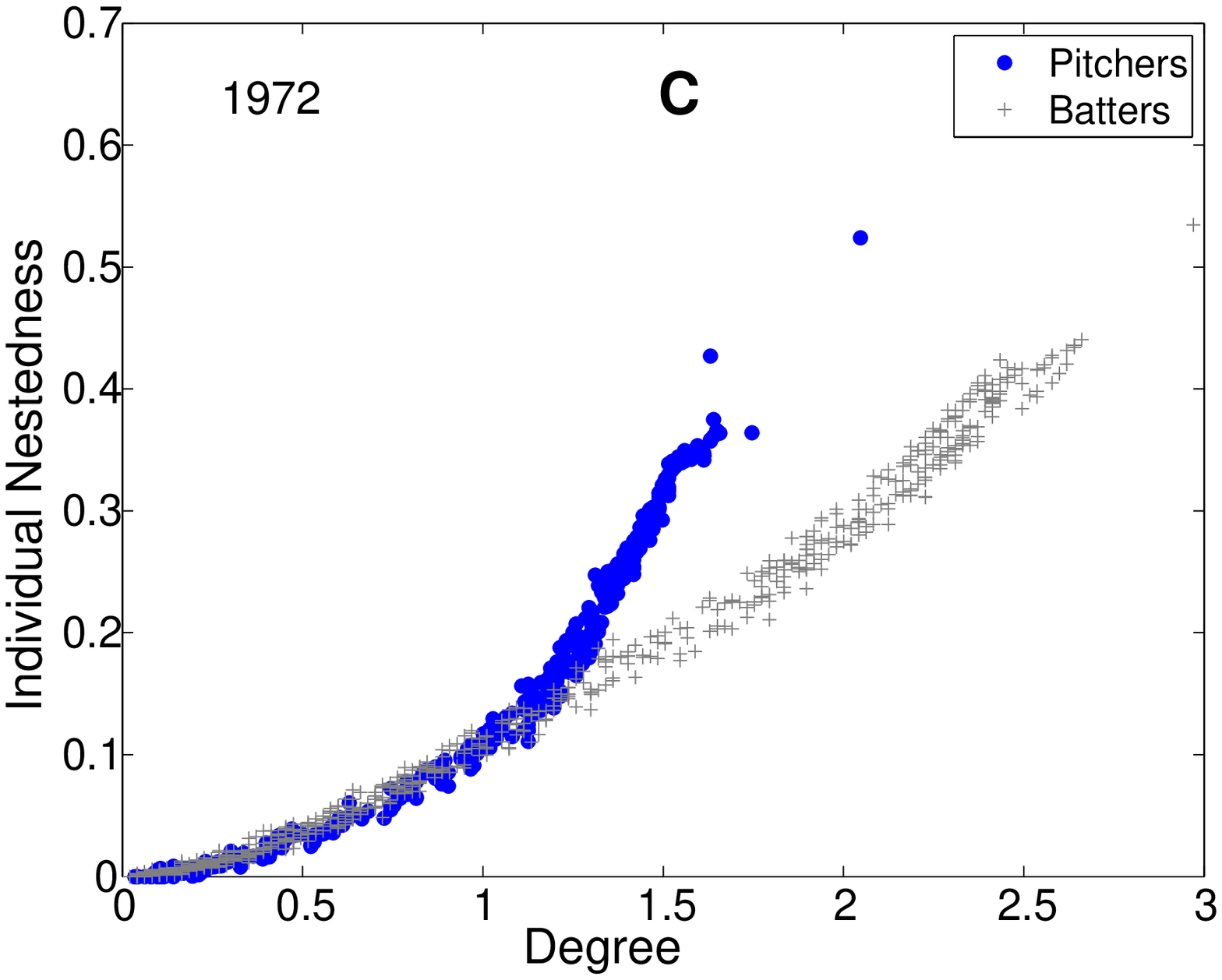}&\includegraphics[width=2.53in]{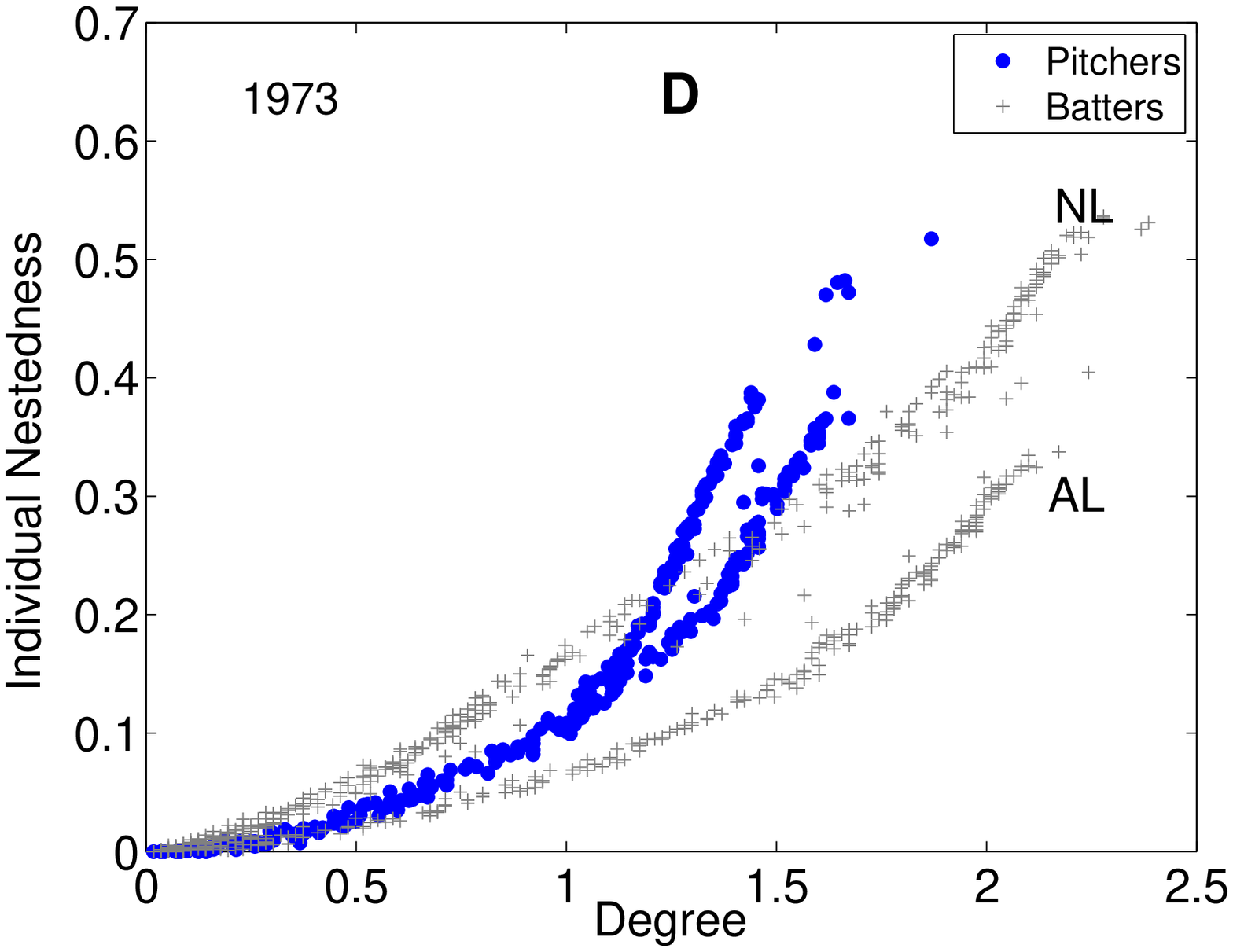}
\end{tabular}
\end{center}
\caption{[Color online] Time Evolution and Summary Statistics of the Baseball Networks. Panel \textbf{A} shows the relation between player degree $k$ and player strength $s$ from 1954 to 2008. The vertical axis gives the value of the exponent $\alpha$ in the power-law relationship $s \sim k^{\alpha}$ (see the discussion in the main text), where we observe that $\alpha$ tends to decrease as a function of time. Shuffling the strengths in the network while keeping the player degrees fixed yields a power-law relationship with $\alpha \approx 1$ for all years. Blue circles denote pitchers and gray crosses denote batters. Each error bar corresponds to one standard deviation. The inset shows on a log-log scale the relationship between degree $k$ and strength $s$ for the 2008 season. Panel \textbf{B} shows the time evolution of the network's nestedness (which we defined using the NODF metric \cite{nodf}). Black circles and red squares represent, respectively, the values for the original data and the standard null model II \cite{nested}.  Each error bar again corresponds to one standard deviation.  Panels \textbf{C} and \textbf{D} show, respectively, the relationship between node degree and individual nestedness for the 1972 and 1973 networks. For comparison purposes, the degree of pitchers and batters are respectively scaled by a multiplicative factor of $P/l$ and $B/l$, where $P$ is the number of pitchers, $B$ is the number of batters, and $l$ is the number of undirected edges in the network. In 1973, the American League introduced the designated hitter rule, which caused a significant change in the structure of subsequent networks.  Between 1954 and 1972, pitchers and batters each collapse onto a single curve.  From 1973 to 2008, however, pitchers and batters each yield two distinct curves, revealing a division between the American league (bottom) and National League (top).
}
\label{fig2}
\end{figure*}

To identify the changes in the organization of baseball networks, we examine the graph properties of single-season networks. The number of distinct opponents per player, given by the distribution of player degree $k_i=\sum_j \hat{A}_{ij}$, follows an exponential distribution for a large range and then has an even faster decay in the tail (see Fig.~\ref{figS1}).  The mean values of the geodesic path length between nodes and of the bipartite clustering coefficient are only somewhat larger than what would be generated by random assemblages (see Appendix \ref{app1}). However, as with mutually-beneficial interactions in ecological networks \cite{asym}, the mutually-antagonistic baseball matchup networks exhibit non-trivial relationships between player degree and player strength $s_i = \sum_j \hat{W}_{ij}$, which represents the total number of opponents of a player (counting multiplicity) \cite{SIREV,handbook}.  As shown in Fig.~\ref{fig2}A, the relation between strength and degree is closely approximated by a power law $s \sim k^{\alpha}$ that starts in 1954 at $\alpha \approx 1.64$ for pitchers and $\alpha \approx 1.41$ for batters but approaches $\alpha \approx 1$ for each by 2008.  The six-decade trend of a decreasing power-law exponent indicates how real-life events such as the increase in the number of baseball teams through league expansion (e.g., in the 1960s, 1977, 1993, and 1998), reorganization (e.g., in 1994, to three divisions in each league instead of two), interleague play (in 1997), and unbalanced schedules (in 2001) have modified the organization of the networks.

An important property mediating the competition dynamics of mutualistic networks in ecology is \textit{nestedness} \cite{Bastolla}.  Although the definition of nestedness may vary, a network is said to be nested when low-degree nodes interact with proper subsets of the interactions of high-degree nodes \cite{nested} (see Fig.~\ref{fig1}).  To calculate the aggregate nestedness in the binary matchup network $\mathbf{A}$, we employed the nestedness metric based on overlap and decreasing fill (NODF) \cite{nodf}, which takes values between $[0,1]$, where $1$ designates a perfectly-nested network (see Appendix \ref{app1}).  Figure~\ref{fig2}B (black circles) shows that single-season baseball networks consistently have nestedness values of approximately $0.28$. This value is slightly but consistently higher than those in randomized versions of the networks with similar distribution of interactions (red squares) \cite{nested}, which we also observe to decrease slightly in time.  In common with bipartite cooperative networks, this confirms that nestedness is a significant feature of these mutually-antagonistic networks.

Although nestedness is defined as a global characteristic of the network, we can also calculate the individual contribution of each node to the overall nestedness \cite{nodf}. Comparing node degrees and individual nestedness (see Appendix \ref{app1}) before 1973, batters and pitchers collapse well onto separate curves (see Fig.~\ref{fig2}C).  Starting in 1973, however, each of these split into two curves (see Fig.~\ref{fig2}D), corresponding to players in the two different leagues: the American League (AL) and the National League (NL). This structural change presumably resulted from the AL's 1973 introduction of the designated hitter (DH), a batter who never fields but bats in place of the team's pitchers (see Fig.~\ref{fig1}), apparently causing the AL to become less nested due to the replacement of low-degree batting pitchers with higher-degree DHs. This suggests, as we discuss below, that the network position of a player might affect his own ranking (while, of course, network position is strongly influenced by a player's longevity and, thus, by his performance).


\section{Biased Random Walkers}\label{sec3}

To compare the performance of players, we rank players by analyzing biased random walkers on the bipartite network $\mathbf{M}$ encoding the outcomes of all mutually-antagonistic interactions between each player pair.  Our method generalizes the technique we previously used for NCAA football teams \cite{notices,monthly}, allowing us to rank players in individual seasons and in the 1954--2008 career network, yielding a quantitative, conceptually-clear method for ranking baseball players that takes a rather different approach from other sabermetric methods used to project player performance such as DiamondMind (which uses Monte Carlo simulations), PECOTA (which uses historical players as a benchmark), and CHONE (which uses regression models) \cite{Pecota,hardball}.

To describe the aggregate interaction $M_{ij}$ between pitcher $i$ and batter $j$, we need to quantify each possible individual pitcher--batter outcome. For simplicity, we focus on the quantity \textit{runs to end of inning} (RUE) \cite{BJ2}, which assigns a value to each possible plate appearance outcome (single, home run, strikeout, etc.) based on the expected number of runs that a team would obtain before the end of that inning, independent of the situational context (see Appendix \ref{app2} for specific values).  Higher numbers indicate larger degrees of success for the batting team.  For each season, we add the RUE from each plate appearance of pitcher $i$ versus batter $j$ to obtain a cumulative RUE for the pair.  Note that any performance index that assigns a value to a specific mutually-antagonistic interaction can be used in place of RUE without changing the rest of our ranking algorithm. We then define the single-season outcome element $M_{ij}$ by the cumulative extent to which the batter's outcome is better ($M_{ij} > 0$) or worse ($M_{ij} < 0$) than the mean outcome over all pitcher--batter matchups that season. When defining the career outcome element $M_{ij}$ for 1954--2008, we account for baseball's modern era offensive inflation \cite{Thorn,BJ2} by summing over individual seasons (i.e., relative to mean outcomes on a per season basis).

We initiate our ranking methodology by considering independent random walkers who each cast a single vote for the player that they believe is the best. Each walker occasionally changes its vote with a probability determined by considering the aggregate outcome of a single pitcher--batter pairing, selected randomly from those involving their favorite player, and by a parameter quantifying the bias of the walker to move towards the winner of the accumulated outcome. A random walker that is considering the outcome described by this matchup is biased towards but not required to choose the pitcher (batter) as the better player if $M_{ij} < 0$ ($M_{ij} > 0$).

The expected rate of change of the number of votes cast for each player in the random walk is quantified by a homogeneous system of linear differential equations $\mathbf{v}' = \mathbf{D}\cdot\mathbf{v}$, where
\begin{equation}
	D_{ij} = \begin{cases}\hat{W}_{ij} + r\hat{M}_{ij}\,, & i\neq j\\
-s_i + r\sum_k\hat{M}_{ik}\,, & i=j\,. \end{cases}
\label{system}
\end{equation}
The long-time average fraction of walkers $\tilde{v}_j$ residing at (i.e., voting for) player $j$ is then found by solving the linear algebraic system $\mathbf{D}\cdot\tilde{\mathbf{v}}=\mathbf{0}$, subject to an additional constraint that $\sum_j \tilde{v}_j = 1$. If the bias parameter $r > 0$, successful players will on average be ranked more highly.  For $r < 0$, the random walker votes will instead tend toward the ``loser'' of individual matchups.

Equation (\ref{system}) gives a general one-parameter system for a biased walker with probabilities that are linear in RUE, but the approach is easily generalized by using other functional forms to map observed plate appearance outcomes (in $\mathbf{M}$) into selection probabilities.  By restricting our attention to a form that is linear in RUE, the interpretation that the off-diagonal components of $\mathbf{D}$ correspond to random walker rate coefficients requires that these components remain non-negative, a preferable state that leads to a number of beneficial properties in the resulting matrix. For example, this allows us to apply the Perron-Frobenius theorem, which guarantees the existence of an equilibrium $\tilde{\mathbf{v}}$ with strictly positive entries (and similarly guarantees the existence of positive solutions in algorithms such as PageRank) \cite{monthly,handbook,Keener,Langville}.  In practice, this requirement is equivalent in the baseball networks to $|r| \lesssim 0.7$, so that the result of a home run in a single plate-appearance matchup (i.e., the case in which a batter faces a pitcher exactly once and hits a home run in that appearance) maintains a small but non-negative chance that a corresponding random walker will still select the pitcher.

However, because the aggregate outcome of most pairings remains close to the mean, the bias in the random walk is small, and the rankings become essentially independent of the bias parameter.  The linear expansion in bias $r$ thereby yields a ranking with no remaining parameters beyond the statistically-selected RUE values, given by $\tilde{\mathbf{v}} = \mathbf{v}^{(0)} + r \mathbf{V} + O(r^2)$\,. Generalizing the similar expansion described in detail in Ref.~\cite{monthly}, the zeroth-order term results in a constant number of votes per player, and the additional contribution at first order is given by the solution of a discrete Poisson equation on the graph:
\begin{equation}
	\sum_j L_{ij} V_j = \frac{4}{n}\sum_j \hat{M}_{ij}\,,
\label{firstorder}
\end{equation}
subject to the neutral charge constraint $\sum_j V_j=0$.  (By analogy with electrostatics, we refer to $V_j$ as the RUE `charge' of node $j$.)  In equation (\ref{firstorder}), $n = P + B$ is the total number of players, $\mathbf{L}=\mathbf{S}-\hat{\mathbf{W}}$ is the graph Laplacian, $\mathbf{S}$ is the diagonal matrix with elements $s_{ii}=\sum_j \hat{W}_{ij}$ (and $s_{ij} = 0$ for $ i \neq j$).  Accordingly, we restrict our attention to the first-order ranking specified by $\mathbf{V}$ and obtained using the solution of equation \eqref{firstorder}.

We tabulate this rank ordering separately for pitchers and batters, for individual seasons and the full career network. We compare the results of the random walker ranking to major baseball awards in Table \ref{tableS1}.  We note that the rankings are highly correlated with the underlying RUE per plate appearance of each player ($r \approx .96$ for 2008 and similar for other seasons), so that the top players in the rankings produced by our method have a strong but imperfect correlation with the lists produced by ranking players according to raw RUE values.  (This similarly holds for any other sabermetric quantity that one might use in place of RUE.)  That is, it matters which players one has faced, and that is codified by the network.  We note, for example, that the differences between random walker rankings and raw RUE rankings appear to appropriately capture the caliber of opponents (for example, pitchers from teams with relatively anemic offenses---such as the 2008 Nationals, Astros, and Reds---have a higher ranking in the random walker ranking, reflecting that they never had the good fortune of going up against their own teams' batters). We also compared the rankings with a leading metric in baseball analysis, ESPN's MLB Player Ratings, which combines ratings from ESPN, Elias, Inside Edge, and \textit{The Baseball Encyclopedia} \cite{espnratings2008}. Of the top 99 players for 2008 who are listed in the Player Ratings, 12 did not meet our threshold for plate appearances. Comparing the random walker results for the remaining 87 players with the Player Ratings yields a $r \approx .5601$ correlation. We thus proceed to study the random walker results for the full career ranking both with confidence that it correlates with methods currently used for single-season analysis and caution that the ranking details do not capture all effects according to current best practices in quantitative baseball analysis \cite{wowyblog}.

The full career ranking allows credible comparisons between players from different eras. Interestingly, considering the rankings restricted to individuals who played in at least 10 seasons during this time (HOF-eligible players), we find that Barry Bonds (batter), Pedro Martinez (pitcher), and Mariano Rivera (relief pitcher) are the best players (in these categories) from 1954 to 2008. We show additional rankings in Table \ref{table1}. We especially note that Albert Belle (29th among batters) is ranked much higher than Jim Rice (115th), suggesting that Belle's hitting performance perhaps merits HOF membership more than that of Rice. Similarly, Bert Blyleven ranks higher not only than current HOF competitors such as Jack Morris and Tommy John but also higher than three HOF pitchers with over 300 wins (Steve Carlton, Phil Niekro, and Don Sutton), which is one traditional benchmark for selecting elite pitchers. Direct comparison with other rank orderings of players across different eras would necessitate restriction to sufficiently similar time periods and is thus beyond the network-science focus of the present study.


\section{Linking Structure to Performance} \label{sec4}

As previously suggested, the network architecture should have important effects on the performance of players. In particular, central players in the network might have a systematic advantage in the rankings relative to those who are not as well connected.  Such structurally-important players (see Table \ref{table1} for examples), who have high values for both betweenness centrality and nestedness, have had long---and usually extremely successful---careers, so it is of significant interest, yet difficult, to gauge the coupled effects on their rank ordering from statistical success versus structural role in the network. In fact, we found no correlation ($r \approx 0.001$) between a player's position---i.e., individual nestedness and betweenness---and his success measured by the fraction of votes received.

\begin{figure*}
\begin{center}
\begin{tabular}{cc}
\includegraphics[width=2.5in]{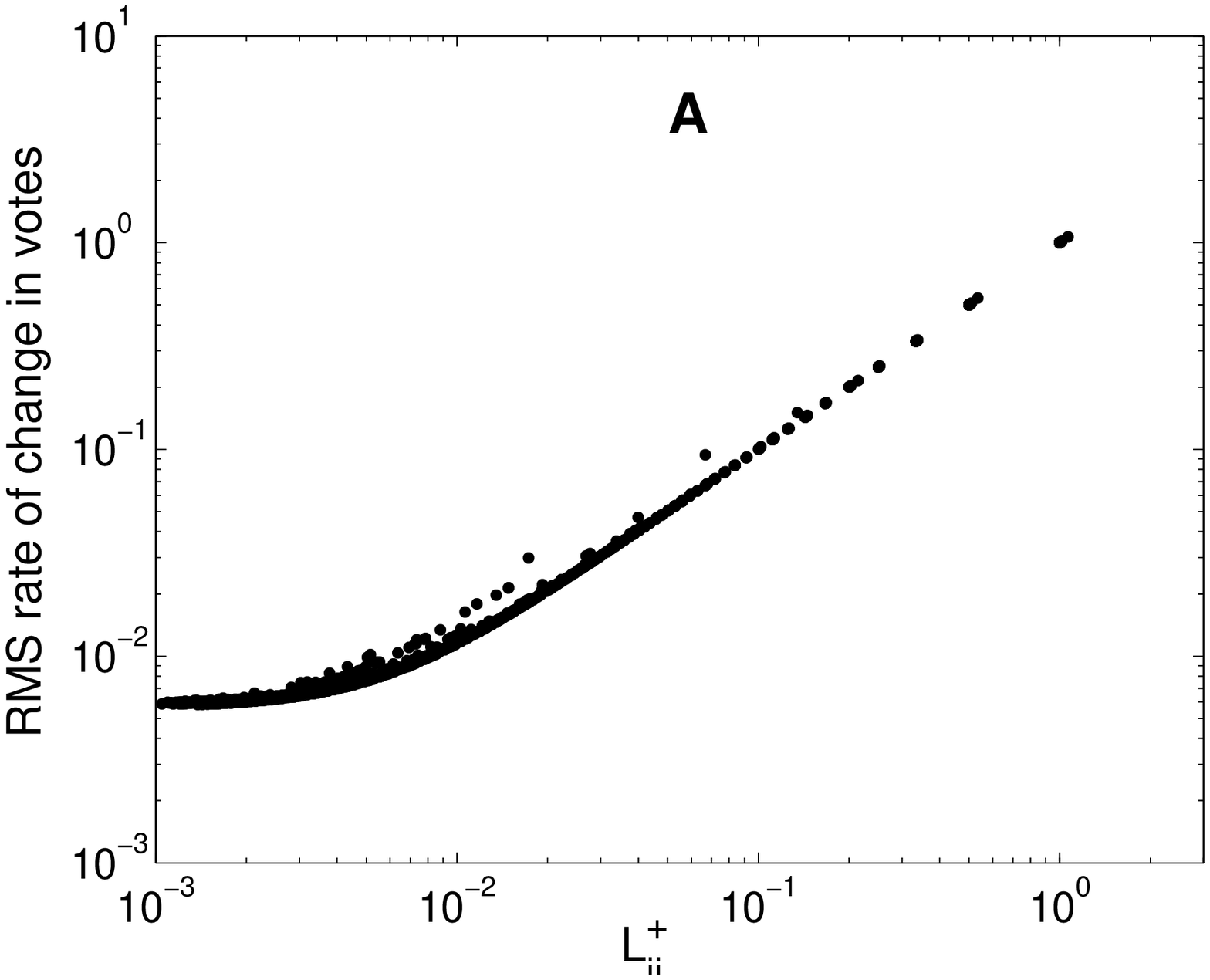}&\includegraphics[width=2.5in]{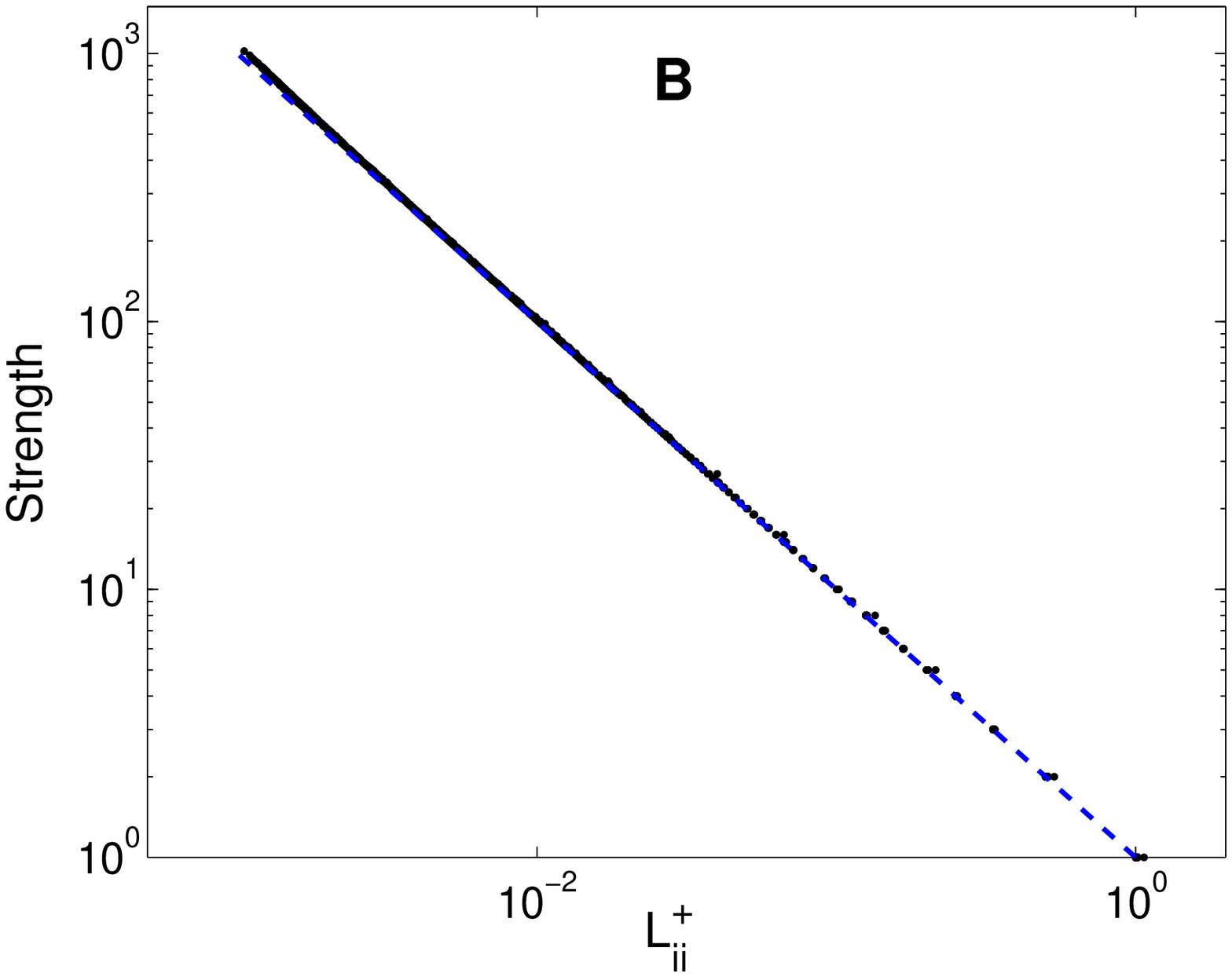}
\end{tabular}
\end{center}
\caption{[Color online] Network Quantities versus Graph Laplacian.  We plot the diagonal elements of the Moore-Penrose pseudo-inverse of the graph Laplacian of the network $L_{ii}^+$ versus (\textbf{A}) the root mean squared change of votes across the network due to the RUE `charge' at each node and (\textbf{B}) node strength.  In each case, we use logarithmic coordinates on both axes. We note in particular the $L^+_{ii}\approx s_i^{-1}$ relationship in panel \textbf{B}.
}
\label{fig3}
\end{figure*}

Hence, we investigate this connection further via the correlation between the sensitivity of rankings to changes in outcomes in individual pitcher--batter pairs, which is formulated using the Moore-Penrose pseudo-inverse of the graph Laplacian. Consider changing the outcome of the single edge that corresponds to the aggregate matchup between players $i$ and $j$.  If we increase the former's aggregate RUE by a unit amount at the expense of the latter, then the total RMS change in votes $\mathbf{V}$ is proportional to the difference between the $i$th and $j$th columns of $\mathbf{L}^+$. This difference yields a node-centric measure of the sensitivities of rankings to individual performances: the constraint $\sum_i L^+_{ij} = 0$ yields that $L^+_{ii}$ (the diagonal element of the graph Laplacian pseudo-inverse), the direct control that player $i$ has over his own ranking, is equal and opposite to the total change his performance directly imposes on the rest of the network. Additionally, as illustrated in Fig.~\ref{fig3}A, the quantity $L^+_{ii}$ is closely related to the RMS changes in votes across the network due to the performance of player $i$.

\begin{figure*}
\begin{center}
\begin{tabular}{cc}
\includegraphics[width=2.5in]{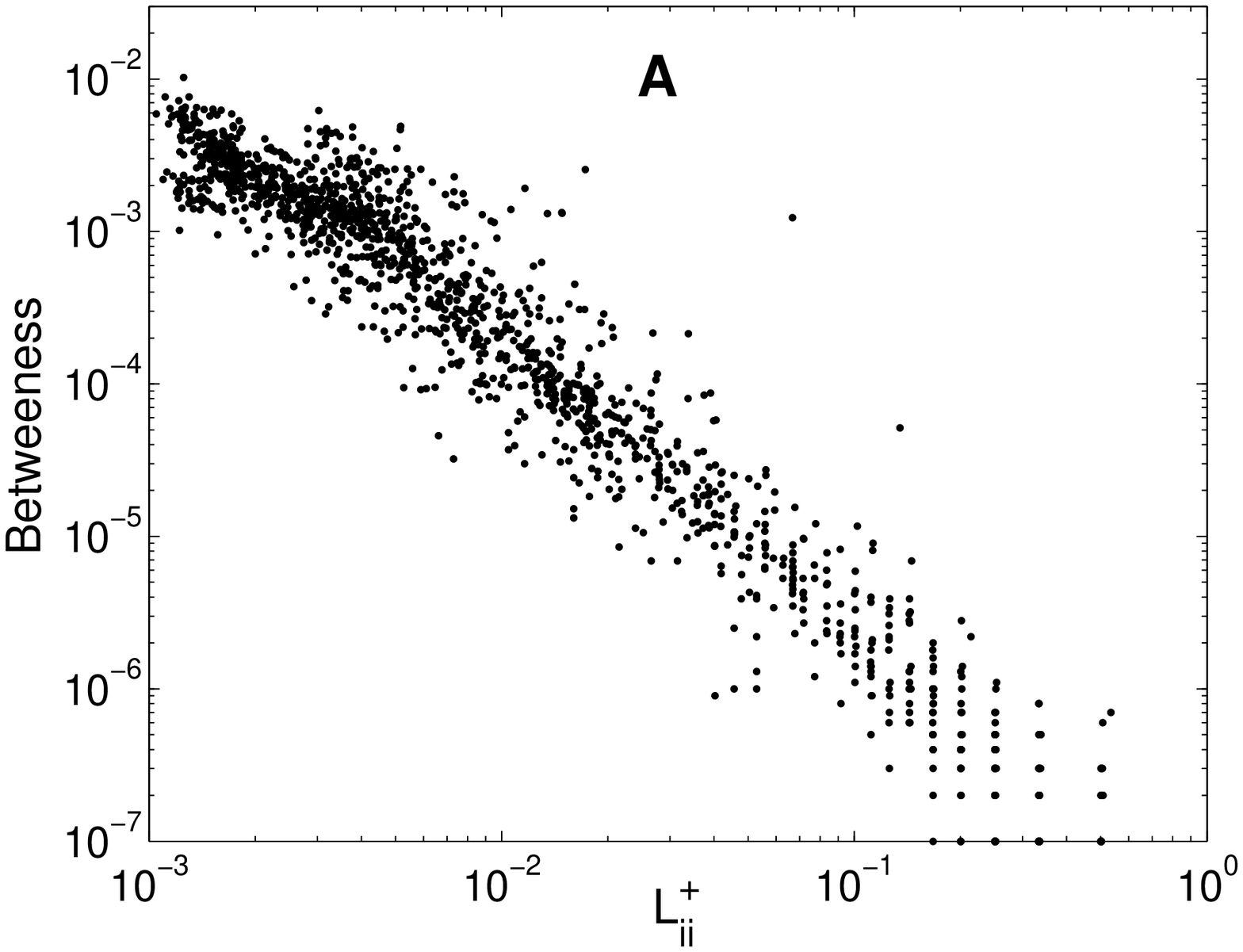}&\includegraphics[width=2.5in]{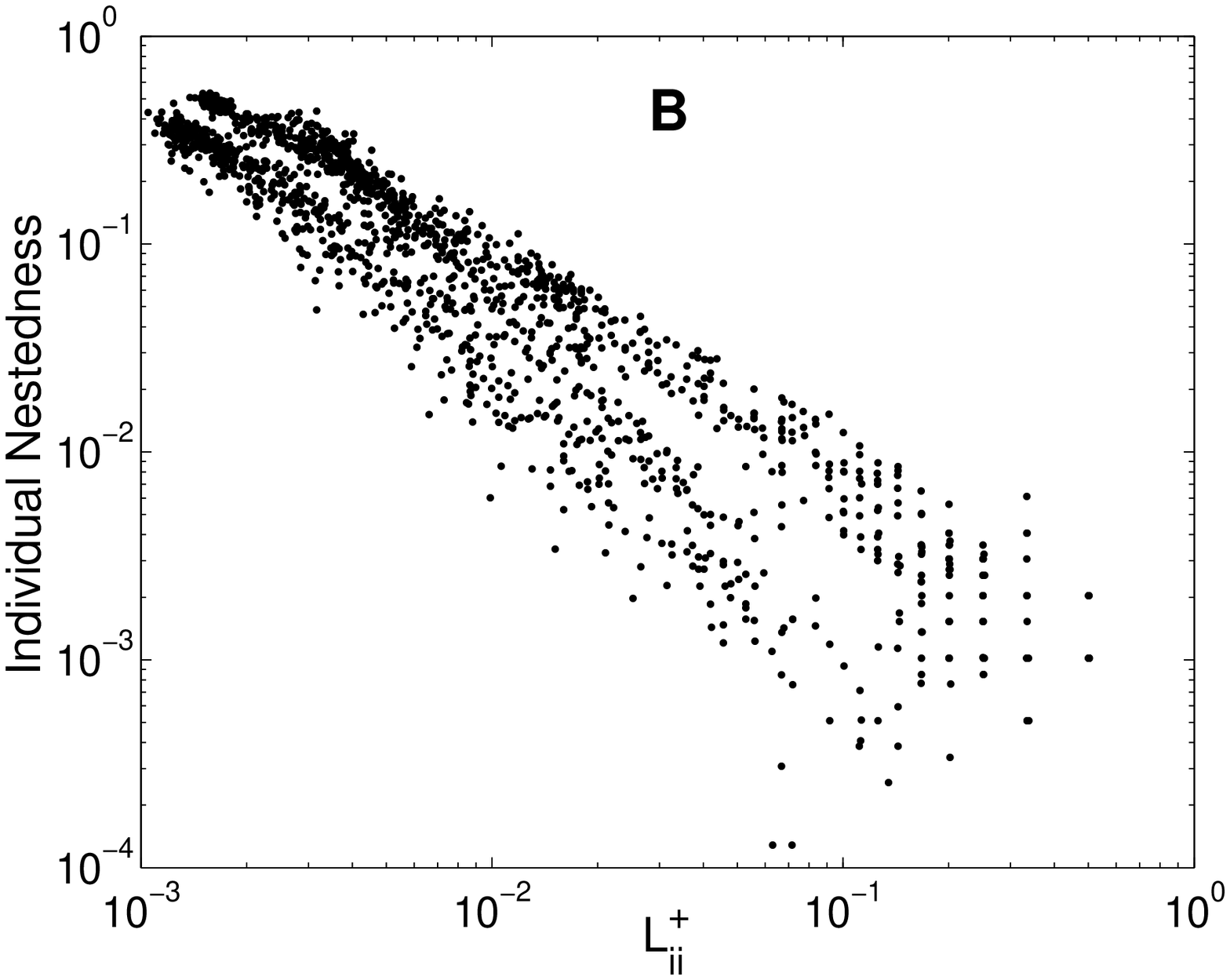}
\end{tabular}
\end{center}
\caption{[Color online] Betweenness and Nestedness versus Graph Laplacian.  We plot the diagonal elements of the Moore-Penrose pseudo-inverse of the graph Laplacian of the network versus (\textbf{A}) node betweenness and (\textbf{B}) individual nestedness.
}
\label{figS2}
\end{figure*}

Noting that the element $L^+_{ii}$ is related to the mean of the commute times between nodes $i$ and $j$ (averaging over all $j$) \cite{aldousfill}, specifically under our constraints, the sum of the commute times $t_{ij} = L^+_{ii} + L^+_{jj} - 2L^+_{ij}$ over $j$ yields a linear function of $L^+_{ii}$.  Consequently, $L^+_{ii}$ provides a node-based measure of the average distance from node $i$ to the rest of the network. This definition of average commute time has similarities with the measures known as information centrality \cite{Stephenson} and random walk centrality \cite{Noh} (though the results of applying the different measures can still be quite different).  The negative relationship between $L^+_{ii}$ and both betweenness centrality and nestedness, which we show in Fig.~\ref{figS2}, thus yields a corresponding negative relationship between the mean commute distance and the betweenness and nestedness of a player.  A player who is highly embedded in the network (i.e., one with high individual nestedness) has a small mean commute distance to the rest of the network, and the ranking of that player is not very sensitive to the outcome of a single matchup. In contrast, a player who is in the periphery of the network (i.e., one with low individual nestedness) typically has a very large mean commute distance to other portions of the graph, and his place in the ranking-ordering is consequently much more sensitive to the results of his individual matchups \footnote{This feature is a sensible one, as players with high nestedness have typically accumulated many more plate appearances, whether as a regular starter in a single-season network or over a long career in the multi-season network, as compared to players further in the periphery of the network.}. This would suggest that players in the AL tend on average to be more prone to changes in their own rankings than players in the NL (see Fig.~\ref{fig2}D).

\begin{figure*}
\begin{center}
\begin{tabular}{cc}
\includegraphics[width=2.5in]{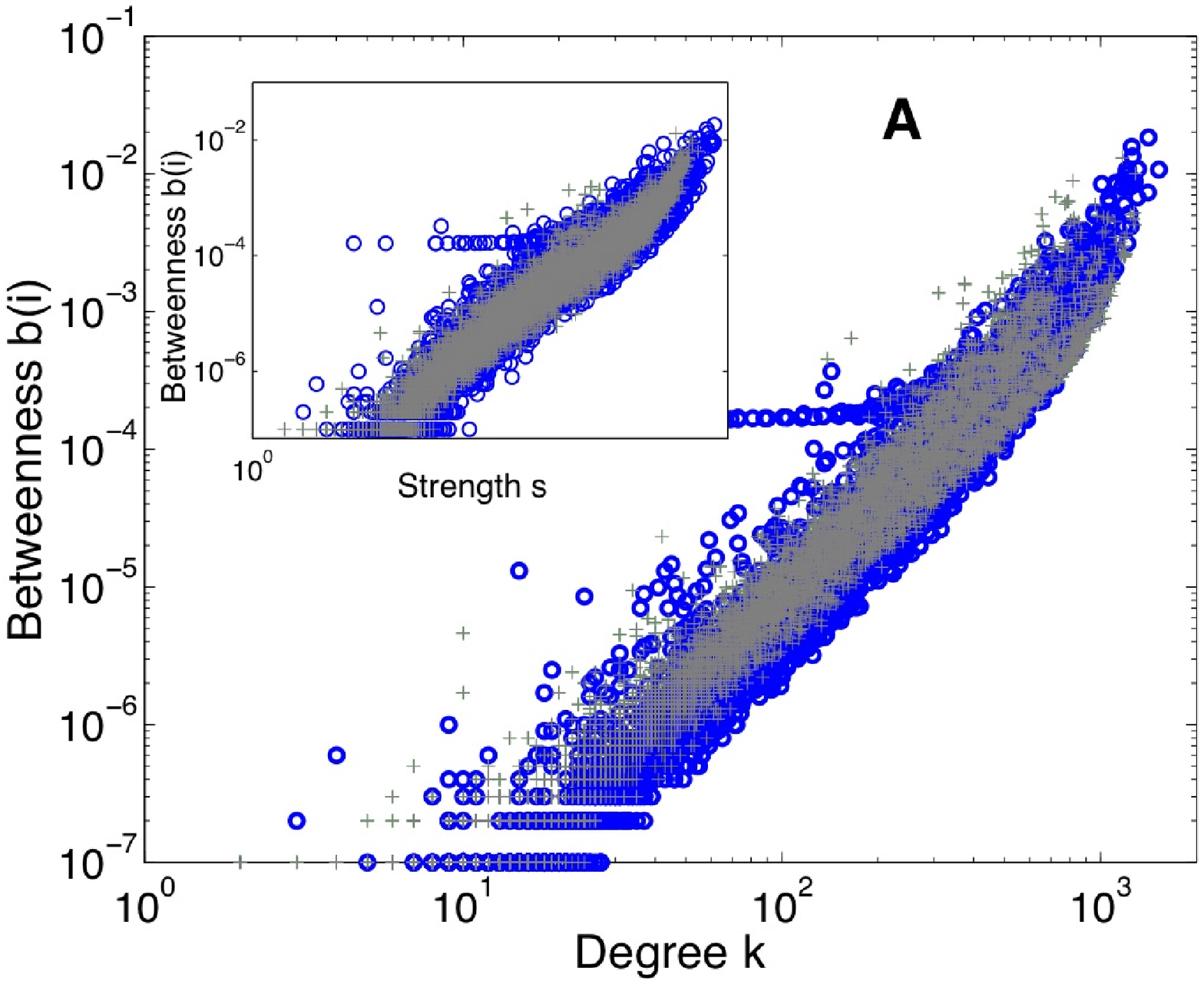}&\includegraphics[width=2.5in]{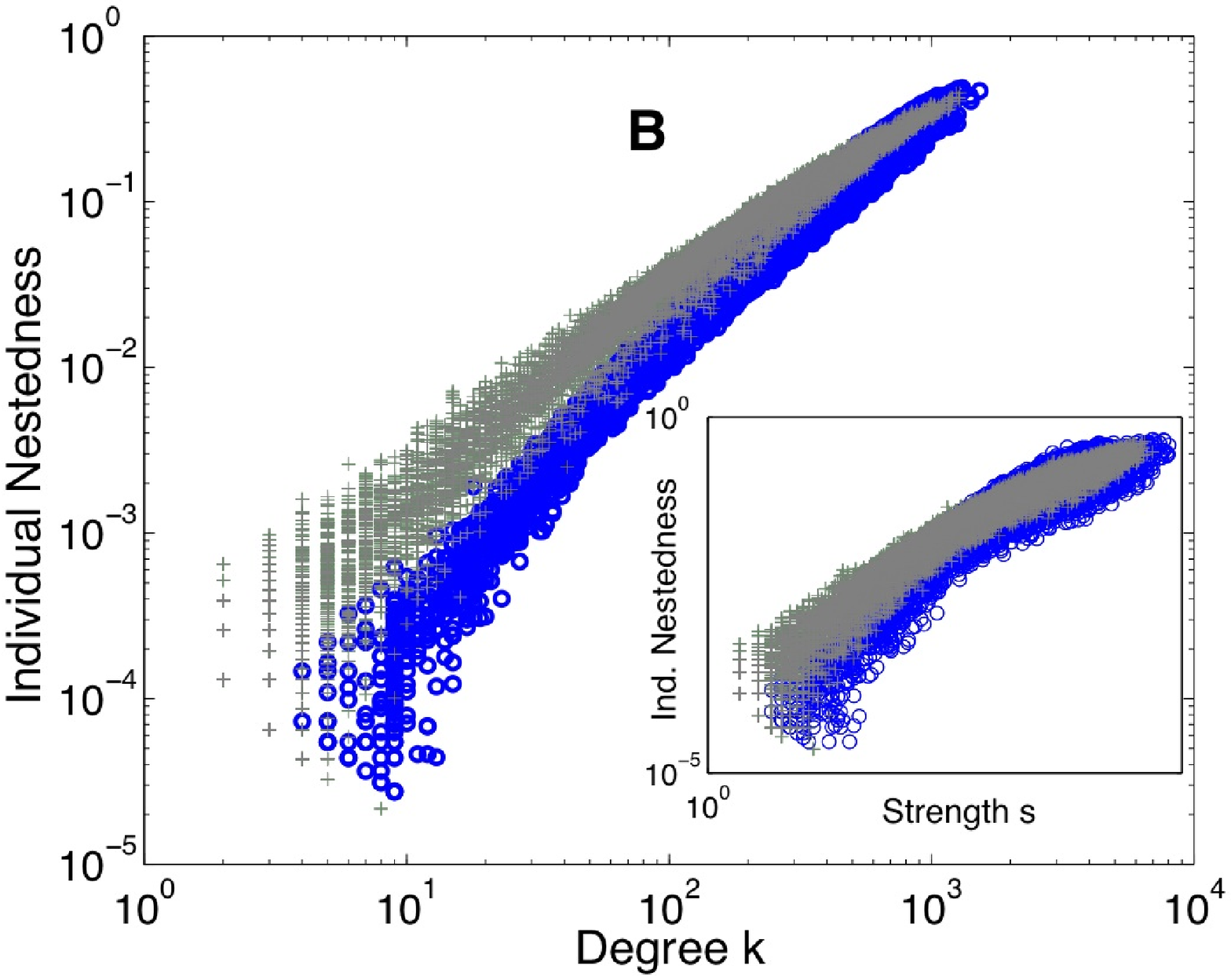}
\end{tabular}
\end{center}
\caption{[Color online] Degree, Strength, Betweenness, and Nestedness. We show a log-log plot of ({\bf A}) player degree $k$ versus betweenness centrality and ({\bf B}) degree versus individual nestedness in the career networks. The insets show the analogous relationships obtained by replacing degree with strength $s$. Pitchers are given by blue dots, and batters are given by gray crosses. Pitchers with betweenness $b \approx 2e^{-4}$ and low degree $k$ tend to be position players who made a few pitching appearances (e.g., Keith Osik), pitchers with short careers (e.g., Wascar Serrano), or recent pitchers with few Major League appearances (e.g., John Van Beschoten, who has split time between the Major Leagues and the Minor Leagues since 2004).
}
\label{figS3}
\end{figure*}

Remarkably, we can make these general notions much more precise, as $L^+_{ii} \approx s_i^{-1}$, where we recall that $s_i$ is the strength of node $i$ (see Fig.~\ref{fig3}B). Some similarities between these quantities is reasonably expected (cf. the role of relaxation times in a similar relationship with random walk centrality in Ref.~\cite{Noh}, which can be quantified by an eigenvalue analysis). This simple relationship belies a stunning organizational principle of this network: The global quantity of average commute time of a node is well-approximated by its strength, a simple local quantity. That is, in the appropriate perturbation analysis to approximate the Laplacian pseudo-inverse, the higher order terms essentially cancel out, contributing little beyond the (zeroth-order) local contribution. We also found a rougher relationship for nestedness and betweenness (see Fig.~\ref{figS3}).

These results have two interesting implications. First, they reveal that the success of well-connected players depends fundamentally on a strong aggregate performance rather than just on their position in the network. Second, they imply that neophyte players would need to face well-connected players if they want to establish a stronger connection to the network and a ranking that is less vulnerable to single matchups.  Similarly, recent research on mutualistic networks in ecology has found that neophyte species experience lower competition pressures by linking to well-connected species \cite{Bastolla}. Our findings on baseball-player rankings suggest the possibility of finding similar competition patterns in mutually-antagonistic interactions in ecological and social networks.


\section{Conclusions}\label{sec5}

Drawing on ideas from network science and ecology, we have analyzed the structure and time-evolution of mutually-antagonistic interaction networks in baseball.  We considered a simple ranking system based on biased random walks on the graphs and used it to compare player performance in individual seasons and across entire careers.  We emphasize that our ranking methodology is overly simplistic, having noted several considerations that one might use to improve it (see, e.g., Appendix B) while maintaining a network framework that accounts for which players each player has faced. We also examined how the player rankings and their sensitivities depend on node-centric network characteristics.

We expect that similar considerations might be useful for developing a better understanding of the interplay between structure and function in a broad class of competitive networks, such as those formed by antigen-antibody interactions, species competition for resources, and company competition for consumers.  Given the motivation from ecology, we are optimistic that this might lead to interesting ecological insights, compensating for the difficulty in collecting data on the regulatory dynamics of mutually-antagonistic networks in ecology---such as the ones formed by parasites and free-living species \cite{parasites}---or assessing the potential performance of invasive species from different environments \cite{Bulleri}.

\begin{acknowledgments}

We thank Eduardo Altmann, Jordi Bascompte, Mariano Beguerisse D\'{i}az, Damian Burch, Justin Howell, Peter Jensen, Tom Maccarone, Ravi Montenegro, Cy Morong, Felix Reed-Tsochas, David Smith, Daniel Stouffer, and Tom Tango for useful comments.  SS did some of his work while a Postdoctoral Fellow at the Oxford University Corporate Reputation Centre and the CABDyN Complexity Centre.  PJM, SP, and TM were funded by the NSF (DMS-0645369) and by start-up funds to PJM provided by the Institute for Advanced Materials, Nanoscience and Technology and the Department of Mathematics at the University of North Carolina at Chapel Hill.   MAP acknowledges a research award (\#220020177) from the James S. McDonnell Foundation. We obtained the baseball network data from \url{retrosheet.org}. The data was obtained free of charge from and is copyrighted by Retrosheet (20 Sunset Rd., Newark, DE 19711).
\end{acknowledgments}

\appendix

\section{Quantities for Bipartite Networks}\label{app1}

Here, we review some important quantities for bipartite networks and discuss their values for the baseball matchup networks.

A clustering coefficient for bipartite networks can be defined by \cite{Zhang}
\begin{equation}
	C_{4,mn}(i) = \frac{q_{imn}}{(k_m-\eta_{imn})+(k_n-\eta_{imn})+q_{imn}}\,, \label{c4}
\end{equation}
where $q_{imn}$ is the number of complete squares involving nodes $i$, $m$, and $n$; $\eta_{imn} = 1+q_{imn}$ enforces the requirement in bipartite graphs that there are no links between nodes of the same population; and we recall that $k_i$ is the degree of node $i$.  Hence, the numerator in (\ref{c4}) gives the actual number of squares and the denominator gives the maximum possible number of possible squares. For the single-sason baseball networks, we calculate the ratio $r_c = \langle C_{4}\rangle/\langle C_{4r}\rangle$ between the mean clustering coefficient $\langle C_{4}\rangle$ summed over all nodes $i$ and the mean clustering coefficient $\langle C_{4r}\rangle$ generated by a randomization of the network that preserves the original degree distribution \cite{Maslov}. We found that baseball networks have average clustering coefficients that are just above that of random networks.  Interestingly, the ratio $r_c$ decreases gradually (and almost monotonically from one season to the next) from $r_c \approx 2.5$ in 1954 to $r_c \approx 1.3$ in 2008.

The geodesic betweenness centrality of nodes over the unweighted network $\hat{\mathbf{A}}$ is defined by \cite{SIREV,Freeman}
\begin{equation}
	b(i) = \sum_{j,k} \frac{\Delta_{j,k}(i)}{d_{j,k}}\,,
\end{equation}
where $\Delta_{j,k}(i)$ is the number of shortest paths between players $j$ and $k$ that pass through player $i$ and $d_{j,k}$ is the total number of shortest paths between players $j$ and $k$. For the single-season baseball networks, we calculate the ratio $r_b=\langle b\rangle /\langle b_r\rangle$ between the mean path length $\langle b \rangle$ summed over all nodes $i$ and the mean path length $\langle b_r \rangle$ generated by a randomization of the network that preserves the degree distribution \cite{Maslov}. As with clustering coefficients, we found that the mean path lengths of baseball networks are only slightly larger than those of random networks, finding in particular that $r_b \in (1,3)$ \cite{Watts}.

Nestedness is an important concept that has been applied to ecological communities, in which species present in sites with low biodiversity are also present in sites with high biodiversity \cite{Atmar}. Although the general notion of nestedness may vary, the concept has nonetheless been employed quite successfully in the analysis of ecological networks \cite{nested}. In a nested network, interactions between two classes of nodes (e.g., plants and animals) are arranged so that low-degree nodes interact with proper subsets of the interactions of high-degree nodes. A nested network contains not only a core of high-degree nodes that interact with each other but also an important set of asymmetric links (i.e., connections between high-degree and low-degree nodes). The importance of nestedness measures is twofold: (1) they give a sense of network organization; and (2) they have significant implications for the stability and robustness of ecological networks \cite{Bastolla,nested}.

To avoid biases in nestedness based on network size (i.e., the number of nodes), degree distribution, and other structural properties, we employ the nestedness calculations introduced recently in Ref.~\cite{nested}. The aggregate nestedness is given by \cite{nodf}
\begin{equation}
	NODF = \frac{\sum_{i,j}N_{i,j} + \sum_{l,m}N_{l,m}}{([P(P-1)/2] + [B(B-1)/2])}\,.
\end{equation}	
For every pair of pitchers ($i$ and $j$), the quantity $N_{i,j}$ is equal to $0$ if $k_i \leq k_j$ and is equal to the fraction of common opponents if $k_i > k_j$.  We also define a similar quantity for every pair of batters ($l$ and $m$).  The nestedness metric takes values between $[0,1]$, where $1$ designates a perfectly-nested network and $0$ indicates a network with no nestedness.

The NODF nestedness also allows us to calculate the individual nestedness of each pitcher (column) or batter (row) using the equation
\begin{equation}
	z(i) = \sum_{j} N_{i,j}/(T - 1)\,,
\end{equation}
where $T = P$ (total number of columns) for pitchers, $T = B$ (total number of rows) for batters and $N_{i,j}$ is calculated as above. In this way, the individual nestedness metric takes values between $[0,1]$, where $1$ designates a perfectly-nested individual and $0$ indicates an individual with no nestedness.

The null model used to compare the empirical nestedness is given by \cite{nested}
\begin{equation}
	q(i,j) = \frac{k_i}{2B} + \frac{k_j}{2P}\,,
\end{equation}
where $q_{i,j}$ is the occupation probability of a pairwise interaction between node $i$ and node $j$, and we recall that $B$ and $P$ are, respectively, the total number of nodes $j$ (batters) and nodes $i$ (pitchers) in the network.  In a bipartite network, $j$ and $i$ represent two different types of nodes, so $q_{i,j}$ is the mean of the occupation probabilities of the row and column. Recent studies have shown that model-generated nestedness values extracted from this null model lower the probability of incorrectly determining an empirical nested structure to be significant \cite{nodf}. For baseball networks, we calculated the standard error---given by $Z = (NODF - \langle NODF\rangle)/\sigma$, where $NODF$ corresponds to the nestedness values of the empirical networks and $\langle NODF\rangle$ and $\sigma$ are, respectively, the average and standard deviations of nestedness values of random replicates generated by the null model. For the baseball networks, we found that $Z > 3$ for all seasons (see Fig.~\ref{fig2}B).


\section{Definition of Runs to End of Inning (RUE)}\label{app2}

To quantify the outcome of each plate appearance, we used the sabermetric quantity \textit{runs to end of inning} (RUE) \cite{BJ2}, which assigns a value to each of the possible outcomes in a plate appearance based on the expected number of runs a team would obtain before the end of that inning following that event, independent of game context.  (RUE can also be adjusted by subtracting the initial run state \cite{wowyblog}.)  Higher numbers indicate larger degrees of success for the batting team.  The batter events (and their associated numerical RUE values) are generic out (0.240), strikeout (0.207), walk (0.845), hit by pitch (0.969), interference (1.132), fielder's choice (0.240), single (1.025), double (1.311), triple (1.616), and home run (1.942).

Note that we are ignoring events such as passed balls and stolen bases that can occur in addition to the above outcomes in a given plate appearance.  This might lead to some undervaluing in the ranking for a small number of position players (such as Tim Raines) that rely on stolen bases.  We also considered the metric known as weighted on base average (wOBA) \cite{Tango}, and note that any metric that assigns a value to a specific plate appearance can be used in place of RUE without changing the rest of our ranking algorithm.  This includes, in particular, popular sabermetric quantities such as win shares and value over replacement player (VORP) \cite{BJ1,BJ2}.  One can also incorporate ideas like ballpark effects into the metric employed at this stage of the algorithm without changing any other part of the method.  Although it would make the methodology more complicated (in contrast to our goals), it is also possible to generalize the algorithm to include more subtle effects such as estimates for when player performance peaks and how it declines over a long career.  Some of the active players in the data set have not yet entered a declining phase in their careers and might have higher rankings now than they will when their careers are over.  We expect that the relatively high rankings of modern players versus ones who retired long ago might also result in part from the increased performance discrepancy between the top players and average players in the present era versus what used to be the case and in part from performing well against the larger number of relatively poor players occupying rosters because of expansion \cite{dictionary}.  Finally, we note that batter--pitcher matchups are not fully random but contain significant correlations (e.g., in a given baseball game, the entire lineup of one team has plate appearances against the other team's starting pitcher) that can be incorporated to generalize the random walker process itself \cite{wowyblog}.

To include the outcome of players who did not have many plate appearances without skewing their rankings via small samples, we separately accumulate the results for all pitchers and batters with fewer than some threshold number of plate appearances $K$ into a single ``replacement pitcher'' and ``replacement batter'' to represent these less prominent players. In the results presented in this paper, we used the threshold $K = 500$ both for single seasons and for the collective outcome matrix.  Note that similar thresholds exist when determining single-season leadership in quantities such as batting average (which requires 3.1 plate appearances per team game, yielding 502 in a 162-game season) and earned run average (1 inning per team game).



\begin{table*}
	\caption{Single-Season Awards and Random Walker Rankings. We show the MVP and CY Young award winners for various years from 1954 to 2008. In parentheses, we give the rank-order of the player within his own category (pitcher or batter) that we obtained using our random walker ranking system applied to the corresponding baseball season. For most of the seasons, there is good agreement between award winners and their random walker ranking.  (Note that the Cy Young award was awarded to a single pitcher---rather than one from each league---until 1967.)}
	\scriptsize
		\begin{tabular}{llll}
& 1954 & 1958 & 1963 \\ \hline
MVP (AL) & Yogi Berra (11th) & Jackie Jensen (8th) & Elston Howard (20th) \\
MVP (NL) & Willie Mays (2nd) & Ernie Banks (6th) & Sandy Koufax (1st) \\
Cy Young (AL) & N/A & Bob Turley (14th) & Sandy Koufax (1st) \\
Cy Young (NL) & N/A & Bob Turley (14th) & Sandy Koufax (1st) \\ \hline				
& 1968 & 1973 & 1978 \\ \hline
MVP (AL) & Denny McLain (4th) & Reggie Jackson (11th) & Jim Rice (3rd) \\
MVP (NL) & Bob Gibson (1st) & Pete Rose (6th) & Dave Parker (1st) \\
Cy Young (AL) & Denny McLain (4th) & Jim Palmer (13th) & Ron Guidry (1st) \\
Cy Young (NL) & Bob Gibson (1st) & Tom Seaver (1st) & Gaylord Perry (30th) \\ \hline				
& 1983 & 1988 & 1993 \\ \hline
MVP (AL) & Cal Ripken Jr. (11th) & Jose Canseco (3rd) & Frank Thomas (3rd)\\
MVP (NL) & Dale Murphy (3rd) & Kirk Gibson (17th) & Barry Bonds (1st) \\
Cy Young (AL) & LaMarr Hoyt (21st) & Frank Viola (24th) & Jack McDowell (17th) \\
Cy Young (NL) & John Denny (14th) & Orel Hershiser (7th) & Greg Maddux (3rd) \\ \hline
& 1998 & 2003 & 2008 \\ \hline
MVP (AL) & Juan Gonzalez (18th) & Alex Rodriguez (7th) & Dustin Pedroia (23rd) \\
MVP (NL) & Sammy Sosa (7th) & Barry Bonds (1st) & Albert Pujols (1st) \\
Cy Young (AL) & Roger Clemens (3rd) & Roy Halladay (15th) & Cliff Lee (8th) \\
Cy Young (NL) & Tom Glavine (10th) & Eric Gagne (8th) & Tim Lincecum (1st) \\ \hline
		\end{tabular}
\label{tableS1}
\end{table*}

\begin{table*}
	\footnotesize
		\begin{tabular}{lllllll}
			Btw(P)&N(P)&R(RP)&R(SP)&Btw(B)&N(B)&R(B) \\ \hline
			Nolan Ryan&Jamie Moyer&Mariano Rivera&Pedro Martinez&Julio Franco&Rickey Henderson&Barry Bonds\\
			Jim Kaat&Roger Clemens&Billy Wagner&Roger Clemens&Rickey Henderson&Barry Bonds&Todd Helton\\
			Tommy John&Greg Maddux&Troy Percival&Roy Halladay&Carl Yastrzemski&Steve Finley&Mickey Mantle\\
			Dennis Eckersley&Mike Morgan&Trevor Hoffman&Curt Schilling&Hank Aaron&Craig Biggio&Manny Ramirez\\
			Jamie Moyer&Randy Johnson&Tom Henke&Sandy Koufax&Pete Rose&Gary Sheffield&Frank Thomas\\
			Greg Maddux&David Wells&B.~J. Ryan&Randy Johnson&Tony Perez&Ken Griffey Jr.&Willie Mays\\
			Charlie Hough&Kenny Rogers&Armando Benitez&John Smoltz&Joe Morgan&Luis Gonzalez&Mark McGwire\\
			Don Sutton&Terry Mulholland&John Wetteland&Mike Mussina&Dave Winfield&Julio Franco&Alex Rodriguez\\
			Phil Niekro&Jose Mesa&Keith Foulke&J.~R. Richard&Ken Griffey Jr.&Jeff Kent&Larry Walker\\
			Roger Clemens&Tom Glavine&Rob Nen&Greg Maddux&Al Kaline&Omar Vizquel&Vladimir Guerrero\\ \hline
		\end{tabular}
					\caption{Player Rankings. Top 10 pitchers (P) and batters (B) according to geodesic node betweenness (Btw), nestedness (N), and random walker ranking (R).  Pitchers are divided into relief pitchers (RP) and starting pitchers (SP). In accordance with HOF eligibility, this table only includes players who played at least 10 seasons between 1954 and 2008. Note that if we consider all players with careers of at least 10 seasons, no matter how many of those seasons occurred between 1954 and 2008, the only change is that Ted Williams becomes the highest-ranking batter.  If we consider all players with at least 8 seasons, the only additional change is that Albert Pujols is ranked just behind Barry Bonds.}
\label{table1}
\end{table*}

\end{document}